\documentclass[a4paper,11pt]{article}
\pdfoutput=1 

\usepackage{jheppub} 
\usepackage[utf8]{inputenc} 
\usepackage[T1]{fontenc} 

\usepackage{feynmp}
\usepackage{amsthm}
\usepackage{bbold}
\usepackage{mmacells}

\newtheorem{observation}{Observation}

\newcommand{\e}{\epsilon}
\newcommand{\bs}[1]{{\boldsymbol{#1}}}
\newcommand{\pd}{\partial}
\newcommand{\T}{\intercal}

\newcommand{\Pexp}{\operatorname{Pexp}}

\title{Symmetric $\e$- and $(\epsilon+1/2)$-forms and quadratic constraints in ``elliptic'' sectors.}

\author{Roman N. Lee}
\affiliation{The Budker Institute of Nuclear Physics,\\ 630090, Novosibirsk}
\emailAdd{r.n.lee@inp.nsk.su}

\abstract{Within the differential equation method for multiloop calculations, we examine the systems irreducible to $\epsilon$-form. We argue that for many cases of such systems it is possible to obtain nontrivial quadratic constraints on the coefficients of $\epsilon$-expansion of their homogeneous solutions. These constraints are the direct consequence of the existence of symmetric $(\epsilon+1/2)$-form of the homogeneous differential system, i.e., the form where the matrix in the right-hand side is symmetric and its $\epsilon$-dependence is localized in the overall factor $(\epsilon+1/2)$. The existence of such a form can be constructively checked by available methods and seems to be common to many irreducible systems, which we demonstrate on several examples. The obtained constraints provide a nontrivial insight on the structure of general solution in the case of the systems irreducible to $\epsilon$-form. For the systems reducible to $\epsilon$-form we also observe the existence of symmetric form and derive the corresponding quadratic constraints.}

\begin{document}
\maketitle
\flushbottom
\section{Introduction}
\label{sec:intro}
Nowadays, multiloop calculations are, in general, much more accessible than before. Thanks to a few powerful techniques, already embodied in the computer codes, we can automatize many stages of multiloop calculations. As a result, many multiloop integrals are readily expressed via generalized polylogarithmic functions. However, it appears that integrals from some families can not be expressed in terms of the polylogarithms,  the simplest example being the two-loop massive sunrise graph.  This annoying circumstance often brings a lot of turbulence to the smooth flow of automatized multiloop calculations. The problem can be localized to the properties of the homogeneous differential equation systems for the master integrals of some specific sectors. Namely, it appears that these systems can not be reduced to $\e$-form, Ref. \cite{Henn2013}, using rational, or even algebraic, transformations, Refs. \cite{Lee2014,Lee2017a}. These irreducible, `elliptic', sectors\footnote{In the present paper we use the term `elliptic' when referring to the sectors in which the differential systems can not be reduced to $\e$-form, unrelated to whether the leading order solutions can be written in terms of elliptic integrals or not.} are, in turn, responsible for the complications in the sectors lying above in the hierarchy\footnote{As usual, by \emph{sector} we mean the integrals with a fixed set of denominators. The latter is defined, of course, up to shift of loop momenta. The hierarchy among the sectors is defined by partial ordering of their sets of denominators with respect to inclusion.}.  A lot of efforts have been made to bring an order in these irreducible cases \cite{Adams2015a,Adams2016,Adams2017,Adams2018,Primo2017,Remiddi2017,Broedel2018b,Adams2018a}. Nevertheless, it looks fair to say that our understanding of the elliptic cases is still far from being complete.

Our dedication in the present paper is to add some insights into the problem of the calculation of multiloop integrals beyond multiple polylogarithms. We observe that the homogeneous system for elliptic cases is often reducible to $(\e+\tfrac12)$-form with symmetric matrix. Relying on this observation and using the dimensional recurrence relations, we derive algebraic constraints on the coefficients of $\e$-expansion of master integrals. We check the validity of these constraints against several known results.

\section{\texorpdfstring{$(\e+\tfrac12)$}{(epsilon+1/2)}-form and constraints for `elliptic' cases}
Our starting point will be the differential equation system for the master integrals of a specific sector:
\begin{equation}
 \pd_x\bs j(x,\e) = M(x,\e) \bs j(x,\e)+\tilde{\bs j}(x,\e)\,,
\end{equation}
where $\bs j(x,\e)$ is a column of master integrals in the chosen sector, $M(x,\e)$ is a matrix rationally depending on both the variable $x$ and dimensional regularization parameter $\e$,  and  $\tilde{\bs j}(x,\e)$ denotes the inhomogeneous term coming from the contribution of the master integrals from subsectors. In the present paper we will concentrate on the homogeneous part of the solution, so we will omit the inhomogeneity  $\tilde{\bs j}$ in what follows. We are mostly interested in the `elliptic' case, when the homogeneous differential system can not be reduced to $\e$-form. Therefore, we will search for generic $\e$-form, where $d=d_0-2\e$, but $d_0$ now is not necessarily equal to $4$. In particular, we make an observation, that differential systems that are `elliptic' for $d_0=4$ are, as a rule, reducible for $d_0=3$ or $d_0=5$.

 Note that, thanks to the dimensional recurrence relations, the differential systems reducible to $\e$-form near some specific $d_0$ are also necessarily reducible near any $d_0+2k$ ($k$ is integer), and vice versa. Therefore, in what follows, we will be sloppy in shifting $\nu$ by an integer, basing sometimes on pure convenience. In particular, in the first and third example in the next section, we use $\nu=1-\e$, while in the second example we have $\nu=2-\e$.
 
To preserve the meaning of $\e$ as (half) a deviation from even dimensionality, $\e=(2n-d)/2$, we will talk about the differential systems in $\e$-form  near $d=2n+1$ as being in the $(\e+\frac12)$-form, so that $d=2n+1-2\tilde{\e}=2n+1-2(\e+\tfrac12)$. So, let us make the following
\begin{observation}
 As a rule, the homogeneous differential system for the master integrals of a specific sector can be reduced at least to one of the two forms: the $\e$-form or  the $(\e+\tfrac12)$-form.
\end{observation}

Strictly speaking, we  can not claim that the above property holds for all irreducible topologies, but we have checked it for several known cases, see, in particular, the examples in the next section.

Suppose now that, using the algorithm of Ref. \cite{Lee2014}, we've managed to find the rational transformation
\begin{equation}
 \bs j(x,\e)=T(x,\e)\bs J(x,\e)
\end{equation}
which casts the differential system to $\mu$-form
\begin{equation}\label{eq:epsilon-form}
 \pd_x\bs J(x,\e) = \mu S(x) \bs J(x,\e)\,,
\end{equation}
where $\mu$ is either $\e$ or $\e+\tfrac12$.

Let us first contemplate how arbitrary is the matrix $S$. In particular, given some rational matrix $S(x)$
can we perform any non-trivial check to decide whether it can or can not appear in the right-hand side of \eqref{eq:epsilon-form}?

To answer this question, let us recall that, in addition to the differential equation \eqref{eq:epsilon-form}, there is also a dimensional recurrence relation (DRR), Ref. \cite{Tarasov1996}, for the column of master integrals, which has the form
\begin{equation}\label{eq:drr}
 \bs j(x,\e-1) = r(x,\e)\bs j(x,\e)\,,
\end{equation}
where $r(x,\e)$ is a matrix rationally depending on $x$ and $\e$. Alternatively, we may write the DRR directly for new master integrals $\bs J$:
\begin{equation}\label{eq:DRR}
 \bs J(x,\e-1) = R(x,\e)\bs J(x,\e)\,,
\end{equation}
where $R(x,\e)=T^{-1}(x,\e-1)r(x,\e)T(x,\e)$ is also  a matrix rationally depending on  $x$ and $\e$.
Now, the compatibility condition of Eqs. \eqref{eq:epsilon-form} and \eqref{eq:DRR} has the form
\begin{equation}\label{eq:compatibility1}
 \partial_x R(x,\e) =  -(1-\mu)S(x)R(x,\e)-\mu R(x,\e) S(x)\,.
\end{equation}
Interpreting Eq. \eqref{eq:compatibility1} as the differential equation for $R$, we see that a non-trivial check for the matrix $S(x)$ to possibly enter the r.h.s. of \eqref{eq:epsilon-form} is the existence of rational solution of \eqref{eq:compatibility1}.

Now we make the following, quite unexpected,
\begin{observation}
 The matrix $S$ in Eq. \eqref{eq:epsilon-form} can be chosen symmetric, $S^{\T}(x)=S(x)$.
\end{observation}

This observation holds for many differential systems that we examined. In addition to the systems reducible to $(\e+1/2)$-form, considered in the next section, we have also checked this property for many systems reducible to $\epsilon$-form\footnote{It worth noting that sometimes the homogeneous equations for the integrals of a specific sector have themselves a block-triangular substructure. Then the symmetricity was observed for each diagonal block.}.

Note that for each specific case it is easy to check whether symmetric form exists. Indeed, if $S^{\T}(x)\neq S(x)$, let us search for the constant transformation $L$, such that $\tilde{S}(x)\equiv L^{-1}S(x)L$ is symmetric, $\tilde{S}^{\T}(x) = \tilde{S}(x)$. Then, multiplying the latter  identity by $L$ and $L^\T$ from the left and right, respectively, we have
\begin{equation}\label{eq: symmetrize}
 \mathcal{L}S^\T(x)-S(x) \mathcal{L}=0\,, 
\end{equation}
where $ \mathcal{L}=LL^\T$.
We are looking for the $x$-independent solution, so, we have a set of linear equations for the elements of the matrix $ \mathcal{L}$ (we should also add equation $\mathcal{L}^\T =\mathcal{L}$). Then, using Cholesky-type decomposition, we can find the matrix $L$ itself.

Now, let's see what interesting consequences the symmetricity of $S$ has. 

We will concentrate on the case $\mu=\e+\tfrac12$, which is relevant for the `elliptic' systems, and comment about reducible cases, where $\mu=\e$, later.
Let $F(x,\e)$ be the fundamental matrix of the differential system \eqref{eq:epsilon-form}, satisfying
\begin{equation}\label{eq:evolution operator}
 \pd_x F(x,\e) = (\e+\tfrac12) S(x)  F(x,\e)\,.
\end{equation}
Now we observe that, if $S^\T(x)=S(x)$, the combination $F^\T(x,-\e) R(x,\e) F(x,\e)$ is independent of $x$, i.e.
\begin{equation}\label{eq:constrFF}
 F^\T(x,-\e) R(x,\e) F(x,\e)=C(\e)\,,
\end{equation}
where $C(\e)$  is a matrix which may depend on $\e$ but not on $x$. Indeed,
\begin{multline}
 \pd_x (F^\T({}-\e) R({}\e) F({}\e))=(\tfrac12-\e) F^\T({}-\e) S^\T{}{}R({}\e) F({}\e)\\
 -F^\T({}-\e) [(\tfrac12-\e)S{}{}R({}\e)+(\tfrac12+\e) R({}\e) S{}{}] F({}\e)+(\tfrac12+\e) F^\T({}-\e) R({}\e) S{}{} F({}\e)\\
 =(\tfrac12-\e) F^\T({}-\e) [S^\T{}{}-S{}{}]R({}\e) F({}\e)=0\,,
\end{multline}
where  we used Eq. \eqref{eq:compatibility1} for $\mu=\e+\tfrac12$, and the last transition is due to symmetricity of $S$. For the sake of better readability we have omitted the argument $x$ in all functions.

It is not difficult to fix the matrix $C(\e)$ in the right-hand side of Eq. \eqref{eq:constrFF}. One only has to take into account that its definition contains some freedom connected with that of $F(x,\e)$. E.g., if we define $F(x,\e)$ as path-ordered exponent with path starting at some regular point $x_0$,
\begin{equation}
 F(x,\e)=U(x,x_0,\e)=\Pexp\left[\left(\e+\tfrac12\right)\int_{x_0}^{x}d \xi \, S(\xi)\right]\,,
\end{equation}
we have
\begin{equation}\label{eq:FRF}
 F^\T(x,-\e) R(x,\e) F(x,\e)=R(x_0,\e)\,.
\end{equation}
Note that the above form of the constraint can be easily established also by writing the formal solution of Eq. \eqref{eq:compatibility1} for $\mu=\e+\tfrac12$ as
\begin{equation}
	R(x,\e)= \Pexp[-(\tfrac12-\e)\intop_{x_0}^{x}d\xi S(\xi)]R(x_0,\e)\overline\Pexp[-(\tfrac12+\e) \intop_{x_0}^{x}d\xi\, S(\xi)] \,.
\end{equation}

Another natural way of using the freedom of definition of $F$ is to fix its asymptotics when $x$ tends to a singular point of $S(x)$. In this case we can calculate the $x$-independent constant simply by taking the corresponding limit.

Once $C(\e)$ is fixed, Eq. \eqref{eq:constrFF} can be readily expanded in $\e$, and gives constraints for each order of $\e$ expansion. The key point here is that the coefficients of $\e$ expansion of $F(x,-\e)$ are the same as those of $F(x,\e)$, up to the alternating sign. More explicitly, if we represent
\begin{equation}
 F(x,\e)=\sum_n F_n(x)\e^n\,,\quad R(x,\e)=\sum_n R_n(x)\e^n\,,\quad
 C(\e)=\sum_n C_n\e^n\,,
\end{equation}
we have the constraints
\begin{equation}
 \sum_{\substack{k,l,m\\k+l+m=n}}(-1)^kF_k^\T(x)R_l(x)F_m(x)=C_n\,.
\end{equation}
Note that quite analogously to Eq. \eqref{eq:constrFF} one can derive the relation
\begin{equation}
 F^\T(x,-\e) R^{\intercal}(x,-\e) F(x,\e)=\tilde C(\e)\,,
\end{equation}
but we observe that, for all differential systems we considered, the matrix $R^{\intercal}(x,-\e)$ was always proportional to $R(x,\e) $ with a factor being independent of $x$. Therefore, the above relation is likely to not give any new constraints additional to those given by Eq. \eqref{eq:constrFF}.

Since the columns of $F$ are the solutions of the homogeneous equation, we can reformulate the constraint \eqref{eq:constrFF} as follows: given any two solutions of the homogeneous equation, $\bs J_1(x,\e)$ and $\bs J_2(x,\e)$, the combination $\bs J_2^\T(x,-\e) R(x,\e) \bs J_1(x,\e)$ is independent of $x$. In particular, this is also valid when we take $\bs J_2=\bs J_1$.

Let us comment about the constraint for the reducible cases. Suppose that $\mu=\e$ in Eq. \eqref{eq:epsilon-form}. Then the fundamental matrix can be written as
\begin{equation}
F(x,\e)=U(x,x_0,\e)=\Pexp\left[\e\int_{x_0}^{x}d \xi \, S(\xi)\right]\,,
\end{equation}
Using the symmetricity of the matrix $S$, it is easy to identify $F^\T(x,-\e)$ with $F^{-1}(x,\e)$, which we can write as a constraint
\begin{equation}
F^\T(x,-\e)F(x,\e)=I\,,
\end{equation}
where $I$ is the identity matrix. We have checked on several examples  that the above identity holds, but does not lead to any new relations between multiple polylogarithms.

\section{Specific examples of `elliptic' cases}

\subsection*{Two-loop sunrise integral}
\begin{figure}[ht]
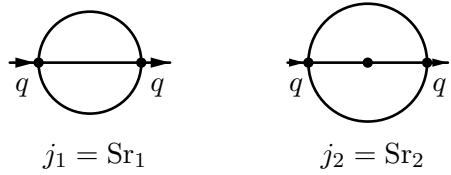

 \centering
  \parbox{3cm}{
   \centering
   \makebox[0pt][l]{\input{sunrise.t1}}\includegraphics{sunrise.1}
   \vspace{1mm}\\
   \centering $j_1=\mathrm{Sr}_1$} \hspace{4mm}
  \parbox{3cm}{
   \centering
   \makebox[0pt][l]{\input{sunrise.t2}}\includegraphics{sunrise.2}
   \vspace{1mm}\\
   \centering $j_2=\mathrm{Sr}_2$}
 \caption{The sunrise topology.\label{fig:sunrise}}
\end{figure}

Let us consider a standard example of the irreducible case --- the sunrise topology with two master integrals depicted in Fig.  \ref{fig:sunrise},
\begin{equation}
 \mathrm{Sr_n}(s=q^2,\e)=\int \frac{d l_1 d l_2}{\left(i\pi^\nu\right)^2\,[l_1^2-1][ l_2^2-1][(q-l_1-l_2)^2-1]^n}\,,
\end{equation}
where the usual $+i0$ prescription is implied, and $\nu=d/2$. We will assume that $d=2-2\e$ ($\nu=1-\e$), i.e., we will consider the expansion near $d=2$. The homogeneous differential system has the form
\begin{equation}\label{eq:srDE}
 \pd_s\bs j(s,\e)
 =\left[
 \begin{array}{cc}
  -\frac{2 \e +1}{s} & -\frac{3}{s}                      \\
  -\frac{(s-3) (2 \e +1) (3 \e +1)}{(s-9) (s-1) s}
                     & -\frac{s^2 \e +s^2+10 s \e -27 \e
  -9}{(s-9) (s-1) s} \\
 \end{array}
 \right]
 \bs j(s,\e)\,,
\end{equation}
where $ \bs j(s,\e)=\bigg(\begin{array}{c}j_1(s,\e)\\j_2(s,\e)\end{array}\bigg)$ is a column of functions.
It is well known that this system can not be reduced to $\e$-form\footnote{The irreducibility can be strictly proved using the criterion of Ref. \cite{Lee2017a}.}.

The differential system \eqref{eq:srDE} has four singular points: $s=0,1,9,\infty$. Therefore, on the general ground, one would not expect the solution to be expressed via hypergeometric functions $_2F_1$. Remarkably, Tarasov in Ref. \cite{Tarasov2006} has found  the general solution of this differential system in terms of the  $\,_2F_1$ using dimensional recurrence relation. Let us write the two homogeneous solutions from Ref. \cite{Tarasov2006} as
\begin{align}
j_1^{(1)}(s,\e) & =\frac{\left(-\frac{s}{(s-1)^2}\right)^{\e }}{s+3} \, _2F_1\left(\frac{1}{3},\frac{2}{3};1-\e
; y\right)\\
j_1^{(2)}(s,\e) & =\frac{(9-s)^{-2 \e } }{s+3} \, _2F_1\left(\frac{1}{3},\frac{2}{3};\e +1;y\right)
\end{align}
Here \[y=\frac{27 (s-1)^2}{(s+3)^3}\,.\]
The corresponding expressions for $j_2^{(1),(2)}$ have the form
\begin{align}
j_2^{(1)}(s,\e) & =\frac{(s-9) s \left(-\frac{s}{(s-1)^2}\right)^{\e } }{9(s-1) (s+3)^2}\, _2F_1\left(\frac{4}{3},\frac{2}{3};1-\e ;y\right)-\frac{(s-3) (1+3 \e)}{9 (s-1)} j_1^{(1)}(s,\e) \\
j_2^{(2)}(s,\e) & =\frac{(s-9) s (9-s)^{-2 \e }}{9 (s-1) (s+3)^2}	\, _2F_1\left(\frac{4}{3},\frac{2}{3};1+\e;y\right)+
\left(\frac{6 \e }{s-9}-\frac{s-3}{9 (s-1)}\right) j_1^{(2)}(s,\e)
\end{align}
Let us now obtain the constraints for the homogeneous solutions and check their validity using the above expressions.
In order to  find the $(\e+\tfrac12)$-form, we pass to the variable $x=\sqrt{s}$ and apply the algorithm of Ref. \cite{Lee2014}. Then we search for the matrix $L$ from \eqref{eq: symmetrize} to obtain the symmetric $(\e+\tfrac12)$-form. This gives us the following transformation
\begin{equation}
\bs j(x^2,\e)= T(x,\e)\bs J(x,\e),\quad
 T(x,\e)=\frac{4^{\e } \Gamma \left(\e
  +\frac{1}{2}\right) \Gamma (3 \e
  +1)}{\sqrt{\pi } \Gamma (\e +1)}\left(
 \begin{array}{cc}
  1 & \frac{\sqrt{3}}{x}          \\
  0 & -\frac{3 \e +1}{\sqrt{3} x} \\
 \end{array}
 \right)\,,
\end{equation}
where  $ \bs J(x,\e)=\bigg(\begin{array}{c}J_1(x,\e)\\J_2(x,\e)\end{array}\bigg)$  are the new functions.
The overall factor $\frac{4^{\e } \Gamma \left(\e+\frac{1}{2}\right) \Gamma (3 \e	+1)}{\sqrt{\pi } \Gamma (\e +1)}$ in the definition of $T(x,\e)$ is not important for the form of the resulting differential system, but simplifies the matrix $R(x,\e)$ entering the dimensional recurrence system. The differential system and dimensional recurrence relations have the forms
\begin{align}
 \partial_x \bs J(x,\e) & =\left(\e+\tfrac12\right)S(x)\bs J(x,\e)\,, \\
 \bs J(x,\e-1)          & =R(x,\e)\bs J(x,\e)\,,
\end{align}
where
\begin{align}
 S(x)                          & =\left[
 \begin{array}{cc}
 -\frac{4 x
 \left(x^2-7\right)}{\left(x^2-9\right)
 \left(x^2-1\right)}           & \frac{4 \sqrt{3}
 \left(x^2-3\right)}{\left(x^2-9\right)
 \left(x^2-1\right)} \\
 \frac{4 \sqrt{3}
 \left(x^2-3\right)}{\left(x^2-9\right)
 \left(x^2-1\right)}           & -\frac{2 \left(x^4+4
 x^2-9\right)}{x \left(x^2-9\right)
 \left(x^2-1\right)} \\
 \end{array}
 \right]\\
 R(x,\e)                       & =\left[
 \begin{array}{cc}
 \left(x^4-30 x^2+45\right) \e & -\sqrt{3}\frac{(x^2-9)
 (x^2-1)+2(x^4-9)
 \e}{x} \\
 \sqrt{3}\frac{ (x^2-9)
 (x^2-1)-2(x^4-9)
 \e}{x}                        & -\frac{3 \left(5 x^4-30
 x^2+9\right) \e }{x^2} \\
 \end{array}
 \right]\,,
\end{align}
Note that $R(x,\e)$ is a linear function of $\e$ with the property $R(x,\e)=-R^{\T}(x,-\e) $.

Let us now write down the constraints. According to the previous section, \linebreak $\bs J^{(a)\T}(x,-\e) R(x,\e) \bs J^{(b)}(x,\e)$ is independent of $x$ (here $\bs J^{(a)}=T^{-1}\bs j^{(a)}$, $a=1,2$).
The constants can be easily fixed by taking the limit $x\to 0$. We have
\begin{align}
	\bs J^{(1)\T}(x,-\e) R(x,\e) \bs J^{(1)}(x,\e)&=-\frac{1}{3} \e  \sin (3 \pi  \e ) \cot (\pi  \e ),\label{eq:J1constr}\\
	\bs J^{(2)\T}(x,-\e) R(x,\e) \bs J^{(2)}(x,\e)&=\frac{1}{3} \e  \sin (3 \pi  \e ) \cot (\pi  \e ),\label{eq:J2constr}\\
	\bs J^{(1)\T}(x,-\e) R(x,\e) \bs J^{(2)}(x,\e)&=\bs J^{(2)\T}(x,-\e) R(x,\e) \bs J^{(1)}(x,\e)=0\,.
\end{align}
The two first constraints result in the following identity
\begin{multline}\label{eq:sunrise_constr}
\, _2F_1\left(\frac{1}{3},\frac{2}{3};1-\e ;y\right) \, _2F_1\left(\frac{1}{3},\frac{2}{3};\e +1;y\right)\\
+\frac{(y-1) }{3 \e }\, _2F_1\left(\frac{2}{3},\frac{4}{3};1-\e ;y\right) \, _2F_1\left(\frac{1}{3},\frac{2}{3};\e +1;y\right)\\
+\frac{(1-y) }{3 \e }\, _2F_1\left(\frac{1}{3},\frac{2}{3};1-\e ;y\right) \, _2F_1\left(\frac{2}{3},\frac{4}{3};\e +1;y\right)=1
\end{multline}
Indeed, this identity is valid, which can be checked independently by first differentiating it and then finding the constant, e.g., via substitution $y\to 0$. 

Let us now examine how the above constraint looks like when expanded in $\e$. 
We have the following identity
\begin{align}\label{eq:FviaH}
\,_{2}F_{1}\left(\tfrac{1}{3},\tfrac{2}{3},1-\e|y\right)&=\sum_{n=0}^{\infty}\e^{n}{H}_{\alpha,\underbrace{1,\ldots,1}_{n}}(y)\,,\\
\,_{2}F_{1}\left(\tfrac{4}{3},\tfrac{2}{3},1-\e|y\right)&=\sum_{n=0}^{\infty}\e^{n}{H}_{\alpha',\underbrace{1,\ldots,1}_{n}}(y)\,,
\end{align}
where
\begin{gather}\label{eq:Hfunc}
{H}_{\beta,\boldsymbol{k}}\left(y\right)=\sum_{j=0}^{\infty}\beta\left(j\right)y^{j}S_{\boldsymbol{k}}\left(j\right)\,,\quad (\beta=\alpha,\alpha')\,,\nonumber \\
\alpha\left(j\right)=\frac{\left(3j\right)!}{3^{3j}\left(j!\right)^{3}}\,,\quad  \alpha'\left(j\right)=\frac{\left(3j+1\right)!}{3^{3j}\left(j!\right)^{3}}.
\end{gather}
Note that ${H}_{\alpha',k}(y)$ can be expressed via ${H}_{\alpha,k}$ and its derivative:
\begin{equation}
	{H}_{\alpha',k}(y)={H}_{\alpha,k}(y)+3y\partial_y{H}_{\alpha,k}(y)\,.
\end{equation}
The function $S_{k_{n}>0,\ldots,k_{1}>0}\left(j\right)$   is defined recursively, as in Ref. \cite{Remiddi:1999ew}:
\begin{equation}
S_{k_{n},\ldots,k_{1}}\left(j\right)=\sum_{i=1}^{j}\frac{1}{i^{k_n}}S_{k_{n-1},\ldots,k_{1}}\left(i\right)\,.
\end{equation}
There is a striking similarity of the definition \eqref{eq:Hfunc} with that of the harmonic polylogarithms \cite{Remiddi:1999ew}.  The only difference is the weight $\alpha(j)$, which, for harmonic polylogarithms, would be $1/j^a$.

\begin{figure}[h]
\begin{mmaCell}[moredefined={SummerTime,TreeSum}]{Input}
<<SummerTime`
Module[\{y = 2/3\}, Timing[
TreeSum[\mmaFrac{(3\,j)! \mmaSup{y}{j}}{\mmaSup{3}{3j} \mmaSup{(j!)}{3}}\mmaFrac{1}{j3}\mmaFrac{1}{j2}\mmaFrac{1}{j1}, \{1,\,\{j1,\,\{j2,\,\{j3,\,\{j\}\}\}\}\}, 500]
]]
\end{mmaCell}
\begin{mmaCell}[]{Output}
\{0.128, 0.4417123360305891\dotfill{}61904952\}
\end{mmaCell}
\caption{\texttt{SummerTime} code for calculation of $H_{\alpha,1,1,1}(y)$ at $y=2/3$.}
\label{fig:program} 
\end{figure} 

From the practical point of view, the advantage of the representation in terms of ${H}_{\alpha,\boldsymbol{k}}(y)$ is that the nested sums in \eqref{eq:Hfunc} have factorized summand, and, therefore, can be calculated without nested loops, e.g., using the  \texttt{SummerTime} package, Refs. \cite{Lee2016d}. To give an impression of how effective is the calculation, we present in Fig. \ref{fig:program} the  \textit{Mathematica} program which calculates $H_{\alpha,1,1,1}(2/3)$ with 500 digits in a fraction of a second.

Substituting Eq. \eqref{eq:FviaH} into Eq. \eqref{eq:sunrise_constr}, we obtain the following set of relations:
\begin{equation}
	\sum_{n=0}^{N}(-)^n H_{\alpha,\bs 1_n}H_{\alpha,\bs 1_{N-n}}
	+[1+(-)^N]y(1-y)\sum_{n=0}^{N+1}(-)^n\left(\partial H_{\alpha,\bs 1_n}\right)H_{\alpha,\bs 1_{N+1-n}}=\delta_{N0}\,.
\end{equation}
For even $N$ the above relation gives a nontrivial constraint for the functions $H_{\alpha,\bs 1_n}$.

\subsection*{Nonplanar two-loop vertex}
\begin{figure}[ht]
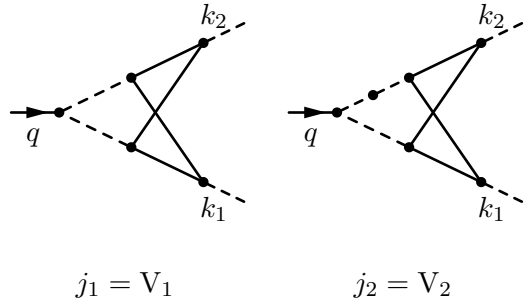

	\centering
		\parbox{3cm}{
			\centering
			\makebox[0pt][l]{\input{vertex.t1}}\includegraphics{vertex.1}
			\vspace{1mm}\\
			\centering $j_1=\mathrm{V}_1$} \hspace{4mm}
		\parbox{3cm}{
			\centering
			\makebox[0pt][l]{\input{vertex.t2}}\includegraphics{vertex.2}
			\vspace{1mm}\\
			\centering $j_2=\mathrm{V}_2$}
	\caption{The two-loop nonplanar  vertex topology, $k_1^2=k_2^2=0$, $q=k_1+k_2$.\label{fig:vertex}}
\end{figure}

Let us consider the second example -- the nonplanar two-loop vertex with massive box as a subgraph, depicted in Fig.  \ref{fig:vertex},
\begin{equation}
\mathrm{V_n}(s=q^2,\e)=\int \frac{d l_1 d l_2\left(i\pi^\nu\right)^{-2}}{[l_1^2]^n\left(q-l_1\right)^2[l_2^2-1][\left(l_2-k_1\right)^2-1]
	[\left(k_2-l_1+l_2\right)^2-1][\left(l_2-l_1\right)^2-1]}\,,
\end{equation}
where the usual $+i0$ prescription is implied, and $\nu=d/2$. This integral has been considered in Ref. \cite{vonManteuffel2017} together with all its subtopologies. For this example we will assume that $d=4-2\e$. The homogeneous differential system has the form
\begin{equation}\label{eq:vDE}
\pd_s\bs j(s,\e)
=\begin{bmatrix}
-\frac{3}{2 s} & -\frac{1}{2} \\
\frac{4 \e +1}{2 s (s+16)} & -\frac{4 s
	\e +7 s+32 \e +80}{2 s (s+16)} \\
\end{bmatrix}
\bs j(s,\e)\,,
\end{equation}
where $ \bs j(s,\e)=\bigg(\begin{array}{c}j_1(s,\e)\\j_2(s,\e)\end{array}\bigg)$ is a column of functions. Note that this system leads to the second-order differential equation for $j_1$, with three singular point, $s=0,16,\infty$. Therefore, we can write the general solution in terms of the hypergeometric functions:
\begin{align}\label{eq:vertexsol}
j_1^{(1)}&=s^{-\e -\frac{3}{2}} \, _2F_1\left(\tfrac{1}{2}-\e ,\tfrac{1}{2}+\e ;1-\e ;-\tfrac{s}{16}\right)\,,\\
j_1^{(2)}&= (s+16)^{-\e } s^{-3/2}\, _2F_1\left(\tfrac{1}{2}-\e ,\tfrac{1}{2}+\e;1+\e;-\tfrac{s}{16}\right)\,.
\end{align}
The corresponding expressions for $j_2^{(1)}$ and $j_2^{(2)}$ easily follow from the equations and will not be presented here.

In order to  find the $(\e+\tfrac12)$-form, we pass to the variable $x$, such that
\begin{equation}
s=\frac{16}{x^2-1}\,.
\end{equation}
Repeating the same steps as in the first example, we end up with the transformation
\begin{equation}
\bs j(s,\e)= T(x,\e)\bs J(x,\e),\quad
T(x,\e)=\Gamma \left(\e +\tfrac{3}{4}\right) \Gamma (\e) \Gamma \left(\e +\tfrac{5}{4}\right)
\begin{bmatrix}
\frac{-8 \left(x^2-1\right)}{4 \e +1} & \frac{8 x \left(x^2-1\right)}{4 \e +1} \\
-\frac12\left(x^2-1\right)^2 & \frac{\left(x^2-1\right)^2}{2x} \\
\end{bmatrix}\,,
\end{equation}
where  $ \bs J(x,\e)=\bigg(\begin{array}{c}J_1(x,\e)\\J_2(x,\e)\end{array}\bigg)$  are the new functions.
The differential system and dimensional recurrence relations have the forms
\begin{align}
\partial_x \bs J(x,\e) & =\left(\e+\tfrac12\right)S(x)\bs J(x,\e)\,, \\
\bs J(x,\e-1)          & =R(x,\e)\bs J(x,\e)\,,
\end{align}
where
\begin{align}
S(x)   =S^\T(\e)                       & =
\begin{bmatrix}
\frac{2 x}{x^2-1} & -\frac{2}{x^2-1} \\
-\frac{2}{x^2-1} & \frac{2}{x(x^2-1)} \\
\end{bmatrix}\\
R(x,\e)     =-R^{\T}(x,-\e)                  & =\begin{bmatrix}
-\frac{\left(3 x^2-1\right) \e }{(x^2-1)^2} & \frac{x \left(2 \e  x^2-x^2+2 \e +1\right)}{2 (x^2-1)^2} \\
\frac{x \left(2 \e  x^2+x^2+2 \e -1\right)}{2 (x^2-1)^2} & \frac{x^2 \left(x^2-3\right) \e }{(x^2-1)^2} \\
\end{bmatrix}\,.
\end{align}

Let us now write down the constraints. We have
\begin{align}
\bs J^{(1)\T}(x,-\e) R(x,\e) \bs J^{(1)}(x,\e)&=\frac{\e ^2 \sin (\pi  \e ) \cos (2 \pi  \e )}{2^{15} \pi ^3},\label{eq:J1constr1}\\
\bs J^{(2)\T}(x,-\e) R(x,\e) \bs J^{(2)}(x,\e)&=-\frac{\e ^2 \sin (\pi  \e ) \cos (2 \pi  \e )}{2^{15} \pi ^3},\label{eq:J2constr1}\\
\bs J^{(1)\T}(x,-\e) R(x,\e) \bs J^{(2)}(x,\e)&=\bs J^{(2)\T}(x,-\e) R(x,\e) \bs J^{(1)}(x,\e)=0\,.
\end{align}
The right-hand sides of these equations are obtained from the limit $x\to 0$.
Again, the two first constraints result in the following identity
\begin{multline}\label{eq:vertex_constr}
\, _2F_1\left(\tfrac{1}{2}-\e ,\e
+\tfrac{1}{2};1-\e ;y\right) \,
_2F_1\left(\tfrac{1}{2}-\e ,\e
+\tfrac{1}{2};\e +1;y\right)\\
\frac{(y-1)
	(1-2 \e) \,
	_2F_1\left(\frac{3}{2}-\e ,\e
	+\frac{1}{2};1-\e ;y\right) \,
	_2F_1\left(\frac{1}{2}-\e ,\e
	+\frac{1}{2};\e +1;y\right)}{2 \e
}\\
-\frac{(y-1) (1-2 \e) \,
	_2F_1\left(\frac{1}{2}-\e ,\e
	+\frac{1}{2};1-\e ;y\right) \,
	_2F_1\left(\frac{3}{2}-\e ,\e
	+\frac{1}{2};\e +1;y\right)}{2 \e
}=1\,.
\end{multline}
Here $y=-\frac{s}{16}$. Note the close resemblance between Eq. \eqref{eq:vertex_constr} and Eq. \eqref{eq:sunrise_constr}.
Eq. \eqref{eq:vertex_constr} can be checked independently by first checking that its derivative is zero it and then finding the constant, e.g., via substitution $y\to 0$. It is possible to examine the $\epsilon$ expansion of the exact hypergeometric solutions \eqref{eq:vertexsol}, with the results being analogous to those of the first example. Namely, one can express any order of expansion in terms of the triangular sums, allowing for the effective high-precision calculation. The summation weights are standard for the multiple hyperlogarithms, except for the first weight, which now has the form
\begin{equation}
\beta\left(j\right)=\left(\frac{\left(2j\right)!}{2^{2j}\left(j!\right)^{2}}\right)^2 \,.
\end{equation}
The exact relation \eqref{eq:vertex_constr} gives nontrivial constraint for each order in $\epsilon$.

\subsection*{Three-loop sunrise integral}

\begin{figure}[ht]
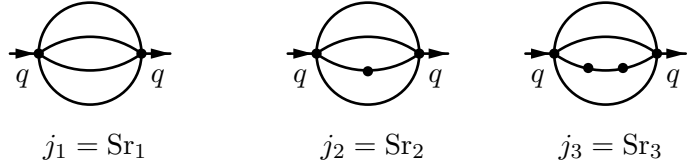

	\centering
		\parbox{3cm}{
			\centering
			\makebox[0pt][l]{\input{sunrise3.t1}}\includegraphics{sunrise3.1}
			\vspace{1mm}\\
			\centering $j_1=\mathrm{Sr}_1$} \hspace{4mm}
		\parbox{3cm}{
			\centering
			\makebox[0pt][l]{\input{sunrise3.t2}}\includegraphics{sunrise3.2}
			\vspace{1mm}\\
			\centering $j_2=\mathrm{Sr}_2$}
		\parbox{3cm}{
			\centering
			\makebox[0pt][l]{\input{sunrise3.t3}}\includegraphics{sunrise3.3}
			\vspace{1mm}\\
			\centering $j_3=\mathrm{Sr}_3$}
	\caption{The three-loop sunrise topology.\label{fig:sunrise3}}
\end{figure}

For the three-loop sunrise topology (also called three-banana), there are three masters which we choose as shown in Fig. \ref{fig:sunrise3}. We will consider the expansion near $d=2$, i.e., define $\e$ via $d=2-2\e$.  For the sake of clear presentation, we do not present the original differential system and recurrence relations.
In order to pass to $(\e+1/2)$-form, we make the transformation
\begin{gather}
\bs j(s=x^2,\e)= T(x,\e)\bs J(x,\e),\\
T(x,\e)=\frac{\left(\frac{16}{3}\right)^{\e } \Gamma \left(2 \e +\frac{1}{2}\right) \Gamma (3 \e +1)}{\Gamma \left(\frac{1}{2}-\e \right) \Gamma (\e +1)}\left(
\begin{array}{ccc}
x & 8 \sqrt{\frac{2}{3}} & \frac{x^2+16}{\sqrt{3} x} \\
0 & -\sqrt{\frac{2}{3}} (4 \e +1) & -\frac{4 (4 \e +1)}{\sqrt{3} x} \\
\frac{(3 \e +1) (4 \e +1)}{2 x} & \frac{(\e +1) (4 \e +1)}{\sqrt{6}} &
\frac{(4 \e +1) (7 \e +5)}{2 \sqrt{3} x} \\
\end{array}
\right)\,.
\end{gather}
The differential system and dimensional recurrence relations for the new function $\bs J(x,\e) $ have the forms
\begin{align}
\partial_x \bs J(x,\e) & =\left(\e+\tfrac12\right)S(x)\bs J(x,\e)\,, \\
\bs J(x,\e-1)          & =R(x,\e)\bs J(x,\e)\,,
\end{align}
where
\begin{align}
S(x)  &=S^{\T}(x)=\left[
\begin{array}{ccc}
-\frac{5 x^2-8}{x \left(x^2-4\right)} & \frac{2 \sqrt{6}}{x^2-4} & -\frac{\sqrt{3} x}{x^2-4} \\
\frac{2 \sqrt{6}}{x^2-4} & -\frac{4 x \left(x^2-10\right)}{\left(x^2-16\right) \left(x^2-4\right)} &
\frac{2 \sqrt{2} \left(5 x^2-32\right)}{\left(x^2-16\right) \left(x^2-4\right)} \\
-\frac{\sqrt{3} x}{x^2-4} & \frac{2 \sqrt{2} \left(5 x^2-32\right)}{\left(x^2-16\right)
	\left(x^2-4\right)} & -\frac{\left(x^2+8\right) \left(3 x^2-16\right)}{x \left(x^2-16\right)
	\left(x^2-4\right)} \\
\end{array}
\right]\\
R(x,\e)&= R_0(x)+\e R_1(x)+\e^2 R_2(x)\\
R_0(x)&=\left(
\begin{array}{ccc}
\frac{1}{4} x^2 \left(x^2-8\right) & \frac{x \left(3 x^2-32\right)}{2 \sqrt{6}} & -\frac{x^4-28 x^2+128}{4 \sqrt{3}} \\
\frac{x \left(3 x^2-32\right)}{2 \sqrt{6}} & \frac{1}{3} \left(x^4-28 x^2+64\right) & -\frac{x \left(5 x^2-16\right)}{6 \sqrt{2}} \\
-\frac{x^4-28 x^2+128}{4 \sqrt{3}} & -\frac{x \left(5 x^2-16\right)}{6 \sqrt{2}} & -\frac{1}{12} x^2 \left(3 x^2-32\right) \\
\end{array}
\right)\\
R_1(x)&=\left(
\begin{array}{ccc}
0 & \frac{x \left(x^2-64\right) \left(x^2-6\right)}{2 \sqrt{6}} & -\frac{5 \left(x^2-8\right) \left(x^2+8\right)}{2 \sqrt{3}} \\
-\frac{x \left(x^2-64\right) \left(x^2-6\right)}{2 \sqrt{6}} & 0 & -\frac{\left(x^2-16\right) \left(x^4+42 x^2-64\right)}{6 \sqrt{2} x} \\
\frac{5 \left(x^2-8\right) \left(x^2+8\right)}{2 \sqrt{3}} & \frac{\left(x^2-16\right) \left(x^4+42 x^2-64\right)}{6 \sqrt{2} x} & 0 \\
\end{array}
\right)\\
R_2(x)&=\left(
\begin{array}{ccc}
-\frac{1}{16} x^2 \left(x^4-104 x^2+832\right) & \frac{x \left(x^2-16\right) \left(x^2+20\right)}{\sqrt{6}} & -\frac{\left(x^2-16\right) \left(x^4-40 x^2-192\right)}{16 \sqrt{3}} \\
\frac{x \left(x^2-16\right) \left(x^2+20\right)}{\sqrt{6}} & \frac{8}{3} \left(x^4-56 x^2+64\right) & \frac{x^6-76 x^4-256 x^2+1024}{3 \sqrt{2} x} \\
-\frac{\left(x^2-16\right) \left(x^4-40 x^2-192\right)}{16 \sqrt{3}} & \frac{x^6-76 x^4-256 x^2+1024}{3 \sqrt{2} x} & -\frac{x^8-8 x^6+3392 x^4-20480 x^2+16384}{48 x^2} \\
\end{array}
\right)\,.
\end{align}
Note the symmetry $R_k=(-)^k R_k^\T$, or, equivalently, $R(x,-\e)=R^\T(x,\e)$.

To the best of our knowledge, there is no closed-form solution of the homogeneous equation for arbitrary $\e$. Remarkably, already for the leading in $\e$ term our constraint is nontrivial.
The explicit expression for this term in terms of the product of elliptic integrals was found in Ref. \cite{Primo2017}. For $j_1(s,\e)$ the results of Ref. \cite{Primo2017} have the form
\begin{equation}
	j_1^{(1)}(s)=\mathrm{K}(\omega_+)\mathrm{K}(\omega_-)\,,\quad
	j_1^{(2)}(s)=\mathrm{K}(\omega_+)\mathrm{K}(1-\omega_-)\,,\quad
	j_1^{(3)}(s)=\mathrm{K}(1-\omega_+)\mathrm{K}(1-\omega_-)\,,	
\end{equation}
where 
\begin{equation}
	\omega_{\pm}=\frac{1}{2}+\frac{s-8}{32} \sqrt{4-s} \pm\frac{s}{32}	\sqrt{16-s} \,,
\end{equation}
and $\mathrm{K}$ is the complete elliptic integral of the first kind. The solutions for $j_{2,3}^{(i)}$ can be deduced from the differential system and are not presented here for the sake of brevity. It is, however, important to note that those solutions contain, in addition to $\mathrm{K}$, the complete elliptic integral of the second kind, $\mathrm{E}$, with argument $\omega_\pm,1-\omega_\pm$.
In order to eliminate square roots,  we introduce a new variable $y$ via
\begin{equation}
	s=-\frac{\left(y^2-9\right)\left(y^2-1\right)}{y^2}\,.
\end{equation}
In terms of this variable we have
\begin{multline}
	\left\{\omega_+,\omega_-,1-\omega_+,1-\omega_-\right\}\\
	=\Bigg\{\frac{(y-1) (y+3)^3}{16
		y^3},\frac{(y-1)^3 (y+3)}{16
		y},
	\frac{(3-y)^3 (y+1)}{16
		y^3},\frac{(3-y) (y+1)^3}{16 y}\Bigg\}\,.
\end{multline}
Expressing $\bs J^{(i)}=T^{-1}\bs j^{(i)}$ and forming the fundamental matrix 
\begin{equation}
	F_0=\Bigg[\bs J^{(1)}\Bigg|\bs J^{(2)}\Bigg|\bs J^{(3)}\Bigg]=\begin{bmatrix}
	J_1^{(1)}& J_1^{(2)} &  J_1^{(2)}\\
	J_2^{(1)}& J_2^{(2)} &  J_2^{(2)}\\
	J_3^{(1)}& J_3^{(2)} &  J_3^{(2)}
	\end{bmatrix} \,,
\end{equation}
we have the constraints
\begin{equation}
F_0^\T R_0 F_0=
\begin{bmatrix}
0 & 0 & -\frac{3 \pi ^2}{4}\\
0 & \frac{9 \pi ^2}{8} & 0 \\
-\frac{3 \pi ^2}{4} & 0 & 0 \\
\end{bmatrix}
\end{equation}
The constant matrix in the right-hand side is obtained from the limit $y\to 0$.
Treating elliptic integrals as independent variables and using Groebner basis approach, we can reduce those constraints to the following system of equations:
\begin{align}\label{eq:constr2}
3 \mathrm{K}_2 \mathrm{K}_3-\mathrm{K}_1
\mathrm{K}_4&=0,\nonumber\\
-6 \mathrm{E}_2 \mathrm{K}_1+2
\mathrm{E}_1 \mathrm{K}_2 y^2-\mathrm{K}_1
\mathrm{K}_2 (y-3) (y+1)&=0,\nonumber\\
18
\mathrm{E}_2 \mathrm{K}_3-2 \mathrm{E}_1
\mathrm{K}_4 y^2+\mathrm{K}_1 \mathrm{K}_4
(y-3) (y+1)&=0,\nonumber\\
-6 \mathrm{E}_4
\mathrm{K}_1+6 \mathrm{E}_3 \mathrm{K}_2
y^2-\mathrm{K}_1 \mathrm{K}_4 (y-1) (y+3)&=0,\nonumber\\
6 \mathrm{E}_4 \mathrm{K}_3-2
\mathrm{E}_3 \mathrm{K}_4 y^2+\mathrm{K}_3
\mathrm{K}_4 (y-1) (y+3)&=0,\nonumber\\
 4y^2\left(3 \mathrm{E}_3 \mathrm{K}_2+
\mathrm{E}_1 \mathrm{K}_4-\mathrm{K}_1
\mathrm{K}_4 \right)^2-9\pi^2&=0\,.
\end{align}
Here
\begin{equation}
\mathrm{K}_{1,2}=\mathrm{K}(\omega_{\pm})\,,\quad \mathrm{K}_{3,4}=\mathrm{K}(1-\omega_{\pm})\,,\quad
\mathrm{E}_{1,2}=\mathrm{E}(\omega_{\pm})\,,\quad \mathrm{E}_{3,4}=\mathrm{E}(1-\omega_{\pm})\,.
\end{equation}
Indeed, we find that the equations \eqref{eq:constr2} hold if we apply the two following known relations
\begin{equation}
	\mathrm{K}_1-\mathrm{K}_2 y = 0\,,\quad 
	3 \mathrm{K}_3-\mathrm{K}_4 y =0\,,
\end{equation}
the identities obtained by the differentiation of the two above, and the Legendre identity
\begin{equation}
	\mathrm{K}_1 \mathrm{E}_3 +\mathrm{E}_1 \mathrm{K}_3- \mathrm{K}_1 \mathrm{K}_3=\frac{\pi}{2}\,.
\end{equation}
So, in this example we see that the obtained constraints can be nontrivial already for the leading in $\e$ order. As to the higher orders in $\e$, we were able to check the constraints in the series expansion over $y$. The approach to construct the coefficients of the generalized power series is described in Ref. \cite{Lee2018a}.

\section{Conclusion}

In the present paper we have obtained nontrivial quadratic constraints on the homogeneous solutions of a few differential systems irreducible to $\epsilon$-form. These constraint appear because these differential systems are reducible to symmetric $(\epsilon+1/2)$-form. Apart from the considered examples, we have checked that similar constraints can be obtained for several other systems irreducible to $\epsilon$-form (the results will be presented elsewhere).

The obtained constraint possibly calls for geometric interpretation. In particular, Eq. \eqref{eq:compatibility1} for $\mu=\epsilon+1/2$ can be written as the `invariance' condition for the tensor field $R(x,\e)$:
\begin{equation}
\nabla_x R(x,\e) = 0\,,
\end{equation}
where 
\begin{equation}
	 \nabla_x =\partial_x+(\tfrac12-\e)S(x)\otimes 1+(\tfrac12+\e)1\otimes S(x)\,.
\end{equation}
Then the constraint \eqref{eq:FRF}
can be viewed as the same invariance condition in integral form. It looks like this invariance should correspond to some properties of the multiloop integrals yet to be discovered. Also, the symmetricity of the matrix in $\e$- and $(\e+1/2)$-form looks very unexpected and deserves a better understanding.

\acknowledgments
I am grateful to Vladimir Smirnov and Matthias Steinhauser for the interest to the work and to the organizers of the workshop ``Taming the complexity of multiloop integrals'', where the results of this paper were first presented. I am especially grateful to Andrei Pomeransky for the interest to the work and many fruitful discussions. This work is supported by the grant of the ``Basis'' foundation for theoretical physics and by RFBR grant 17-02-00830.

\bibliographystyle{JHEP}

\providecommand{\href}[2]{#2}\begingroup\raggedright\endgroup

\end{document}